\documentclass[8pt,osajnl2,onecolumn,superscriptaddress]{revtex4} 

\usepackage{amsmath}
\usepackage{graphicx}

\begin{document}

\title{Image Transmission Through an Opaque Material}

\affiliation{Institut Langevin, ESPCI ParisTech, CNRS UMR 7587, Universit\'es Paris 6 \& 7, INSERM, ESPCI, 10 rue Vauquelin, 75005 Paris, France.}

\author{S\'ebastien Popoff, Geoffroy Lerosey, Mathias Fink, Albert Claude Boccara, Sylvain Gigan}

\maketitle

\textbf{
Optical imaging relies on the ability to illuminate an object, collect and analyze the light it scatters or transmits. Propagation through complex media such as biological tissues was so far believed to degrade the attainable depth as well as the resolution for imaging  because of multiple scattering. This is why such media are usually considered opaque.  Very recently, we have proven that it is possible to measure the complex mesoscopic optical transmission channels that allows light to traverse through such an opaque medium. Here we show that we can optimally exploit those channels to coherently transmit and recover with a high fidelity an arbitrary image, independently of the complexity of the propagation.}

In a classical optical system, the propagation of a complex field from one plane to another is well understood, be it by Fresnel or Fraunhofer diffraction theory, or ray-tracing for more complex cases~\cite{goodman2005introduction}. However, all these approaches break down when multiple scattering occurs~\cite{sebbah2001introduction}. A medium where light is  scattered many times mixes in a seemingly random way all input k-vectors, and is usually considered opaque. Until recently, scattering has always been considered as noise~\cite{dylov2010nonlinear} and most imaging techniques in turbid media rely on ballistic photons only~\cite{wang1991ballistic,shiratori1998photorefractive} which prevents the study of thick scattering samples. Following works in acoustics~\cite{fink1995robust}, recent experiments have demonstrated that multiply scattered light can nonetheless be harnessed, thanks to wavefront control~\cite{yaqoob2008optical, vellekoop2007focusing,cizmar2010situ}, and even put to profit to surpass what one can achieve within a homogenous medium in terms of focusing~\cite{vellekoop2010exploiting}. 

\begin{figure}[ht]
\center
\includegraphics[width=0.6\textwidth]{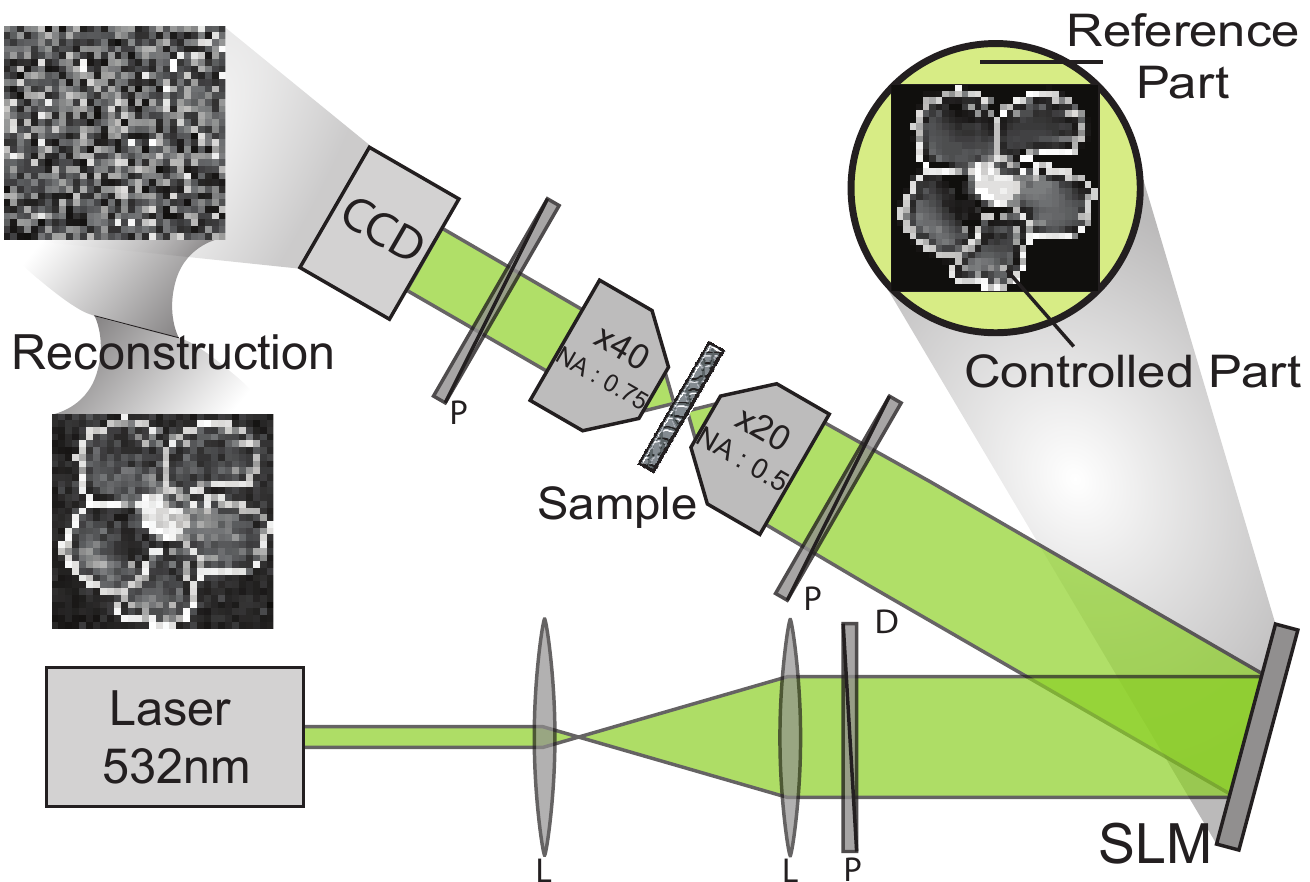}
\caption{\textbf{Experimental setup.} A 532 nm laser is expanded and reflected off a spatial light modulator (SLM) . The laser beam is  phase-modulated then focused on the multiple-scattering sample and the output intensity speckle pattern is imaged by a CCD-camera. L, lens. P, polarizer. D, diaphragm. The object to image is synthetized directly by the SLM, and reconstructed from the output speckle thanks to the transmission matrix.} 
\label{Setup}
\end{figure}

In our experiment (see figure \ref{Setup}), we illuminate with a laser an object (displayed via a spatial light modulator, or SLM), and recover its image on a CCD camera, after propagation through a thick opaque sample.  As expected, we measure on the camera a speckle, that bears no resemblance to the original image. This speckle is the result of multiple scattering and interferences in the sample. 

Although it can be described on average by diffusion equation or Monte Carlo simulation~\cite{sassaroli1998monte}, the propagation through a real linear multiple scattering medium is too complex to be described by classical means. Nonetheless, multiple scattering is deterministic and information is not lost. In other terms, the measured pattern  on the CCD is the result of the transmission of light through a large number of very complicated optical channels, each of them with a given complex transmission. Here, we study the inverse problem of the reconstruction of an arbitrary image, and show that it is possible to recover it through the opaque medium. A prerequisite is however to measure the so-called transmission matrix (TM) of our optical system. 

We define the mesoscopic transmission matrix (TM) of an optical system for a given wavelength as the matrix $K$ of the complex coefficients $k_{mn}$ connecting the optical field (in amplitude and phase) in the $m^{th}$ of $M$ output free mode to the one in the $n^{th}$ of $N$ input free mode. Thus, the projection $E^{out}_m$ of the outgoing optical field on the $m^{th}$ free mode is given by $E^{out}_m = \sum_n{k_{mn}E^{in}_n}$ where $E^{in}_n$ is the complex amplitude of the optical field in the $n^{th}$ incoming free mode.  In essence, the TM gives the relationship between input and output pixels, notwithstanding the complexity of the propagation, as long as the medium is stable. A Singular Value Decomposition (SVD) of the TM gives the input and output eigenmodes of the system and singular values are the amplitude transmission of these modes.

Inspired by various works in acoustics~\cite{tanter2000time,montaldo2004} and electromagnetism~\cite{lerosey2007focusing}, we demonstrated in~\cite{popoff2009measuring} that it is possible to measure the TM of a linear optical system that comprises a multiple scattering medium. In a nutshell, we send several different wavefronts with the SLM, record the results on the CCD, and deduce the TM using phase-shifting interferometry. The singular value distribution of a TM of a homogeneous zone of the sample follows the quarter-circle law (i.e. there is no peculiar input/output correlation~\cite{marcenko1967distribution}) which indicates that light propagation is in the multiple scattering regime with virtually no ballistic photons left.

Using this technique, we have access to $K_{obs} = K\times S_{ref}$, where $S_{ref}$ is a diagonal matrix due to a static reference speckle. The input and output modes are the SLM and the CCD pixels respectively. The measured matrix $K_{obs}$ is sufficient to recover an input image. This TM measurement takes a few minutes, and the system is stationary well over this time. Once the matrix is measured, we generate an amplitude object $E_{obj}$ by subtracting two phase objects (see Methods for details). A realization takes a few hundred ms, limited only by the speed of the SLM.

Here, our aim is to use the TM to reconstruct an arbitrary image through the scattering sample: we need to estimate the initial input  $E_{obj}$ from the output amplitude speckle $E_{out}$.  This problem consists in using an appropriate combination of the medium channels and therefore using a weighting of singular modes/singular values of the TM matched to the noise and to the transmitted image. Noises of different origins (laser fluctuation, CCD readout noise, residual amplitude modulation) degrade the fidelity of the TM measurement. It is the exact analog of Multiple-Input Multiple-Output (MIMO) information transmission in complex environment that have been studied in the past few years in wireless communications~\cite{introPaulraj}. This inverse problem also bears some similarities to optical tomography~\cite{arridge1999optical,corlu2007three}, albeit in a coherent regime~\cite{maire2009experimental}. 

 There are two straightforward options. (i) Without noise, a perfect image transmission can be performed by the use of the inverse matrix (or pseudo-inverse matrix for any input/output pixels ratio) since $K_{obs}^{-1}K_{obs} = I$ where I is the identity matrix. Unfortunately, this operator is very unstable in presence of noise. Singular values of $K_{obs}^{-1}$ are the inverse ones of $K_{obs}$, thus singular values of $K_{obs}$ below noise level result in strong and aberrant contributions. The reconstructed image can hence be unrelated with the input. (ii) In a general case, another possible operator for image transmission is the Time Reversal operator. This operator is known to be stable regarding noise level since it takes advantage of the strong singular values to maximize energy transmission~\cite{tanter2000time}. Its monochromatic counterpart is phase conjugation (classically used to compensate dispersion in optics~\cite{yariv1979compensation}) and is performed using $K_{obs}^{\dag}$. $K_{obs}^{\dag}K_{obs}$ has a strong diagonal but the rest of it is not null, which implies that the fidelity of the reconstruction rapidly decreases with the complexity of the image to transmit~\cite{derode2001scattering}.
A more general approach is to use a Mean Square Optimized operator (MSO), that we note $W$. This operator minimizes transmission errors~\cite{tikhonov1963solution,introPaulraj}, estimated by the expected value $E\left\{\left[W.E^{out}-E^{in}\right]\left[W.E^{out}-E^{in}\right]^{\dag}\right\}$.  For an experimental noise of standard deviation $\sigma$ on the output pixels, $W$ reads :

\begin{equation}
W = \left[K_{obs}^{\dag}.K_{obs} + \sigma.I \right]^{-1}K_{obs}^{\dag}.
\end{equation}

Without noise, $W$ reduces to the inverse matrix $K_{obs}^{-1}$, which is optimal in this configuration, while for a very high noise level it becomes proportional to the transpose conjugate matrix $K_{obs}^{\dag}$, the phase conjugation operator. It is important to note that $\sigma$ has the same dimension as $K_{obs}^{\dag}.K_{obs}$ and thus has to be compared with the square of the singular values of $K_{obs}$. Because of this experimental noise the reconstruction is imperfect. We estimate the reconstruction fidelity by computing the correlation between the image and the object.

 A general principle is that the reconstruction noise can be lowered by increasing the number of degree of freedom ($N_{DOF}$) that we measure and control. For a given object corresponding to $N$ input pixels, we investigated two possibilities: averaging over disorder realizations and increasing the number of output modes $M$.
 
A possible way to average over disorder is to illuminate the object with different wavefronts. It is formally equivalent to transmitting the same image through different channels as if the image propagated through different realizations of disorder. To that end, we use different combinations of random phase masks to generate the same 'virtual object' (see methods). We use this technique to virtually increase $N_{DOF}$, and we average the results to lower the reconstruction noise. It is the mochromatic equivalent of using broadband signals, which takes advantage of temporal degree of freedom~\cite{lemoult2009}. We show in figure 
2 the results for the image transmission of a gray-scale 32 by 32 pixels pattern, and detected on a 32 by 32 pixels region on the CCD. We tested MSO at different noise level for one realization and for averaging over 40 'virtual realizations' with random phase masks. To find the optimal MSO operator, we numerically compute the optimal $\sigma$ that maximizes the image reconstruction, hence obtaining an estimation of the experimental noise level. A simple inverse filtering does not allow image reconstruction, even with averaging, while phase conjugation converges toward 75\% correlated image. In contrast, optimal MSO, allowed a 94\% correlation for 40 averaging (and a modest 34\% correlation in one realization). In addition, Optimal MSO is very robust to the presence of ballistic contributions that strongly hinder reconstruction in phase conjugation (see discussion).

\begin{figure}[ht]
\center
\includegraphics[width=1\textwidth]{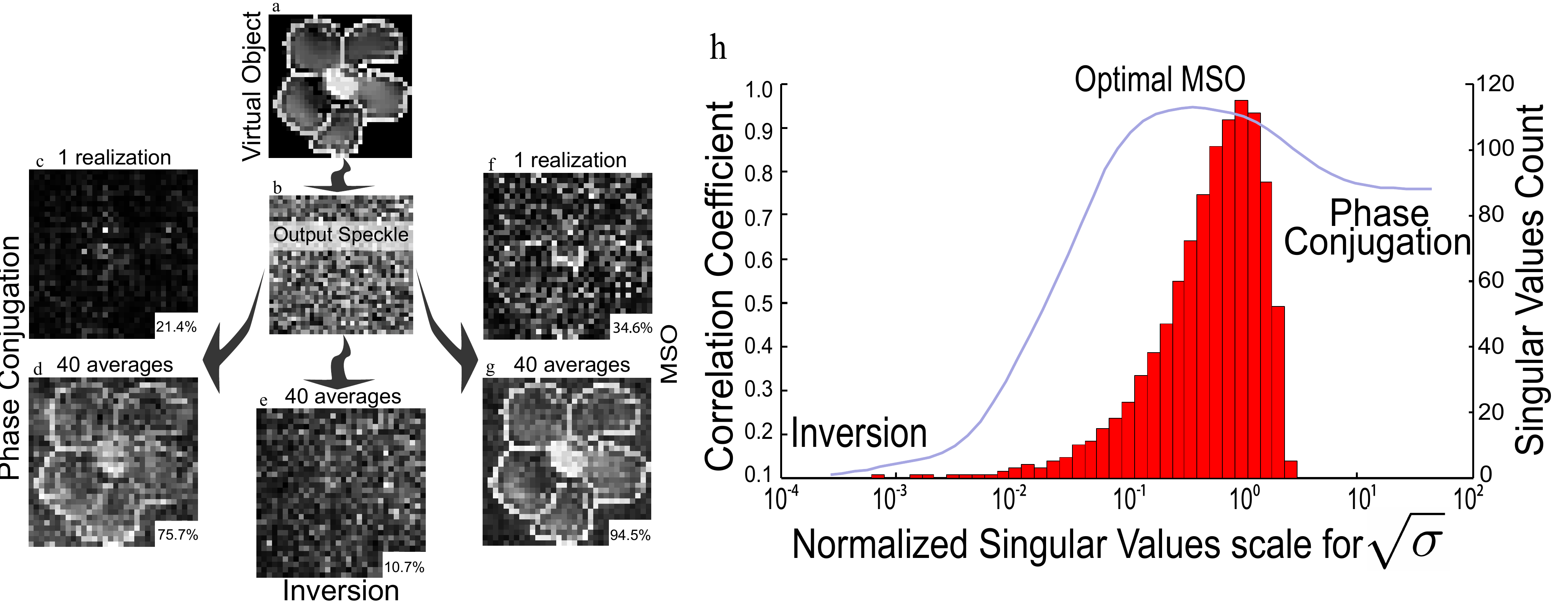}
\caption{\textbf{Comparisons of the reconstruction methods.} \textbf{a.} initial gray scale object and \textbf{b.} a typical output speckle figure after the opaque medium. \textbf{c.} and \textbf{f.} are experimental images obtained with one realization using respectively phase conjugation and MSO operator, \textbf{d.}, \textbf{e.} and \textbf{g.} are experimental images averaging over 40 'virtual realizations' using respectively inverse matrix, phase conjugation and MSO operator. Values in insets are the correlation with the object \textbf{a.}. \textbf{h.} Correlation coefficient between $E_{img}$ and $E_{obj}$ as a function of $\sqrt{\sigma}$ (line) and singular value distribution of $K_{obs}$ (bars). Results are obtained averaging over 100 'virtual realizations' of disorder and both $\sqrt{\sigma}$ and singular values share the same scale on the abscissa axis.} 
\label{Results}
\end{figure}

The second approach in order to add degrees of freedom is to increase the number $M$ of independent pixels recorded on the CCD. In contrast with focusing experiments where the quality of the output image depends on the number of input modes $N$~\cite{vellekoop2007focusing}, the quality of image reconstruction depends on the number of output modes $M$. An important advantage is that the limiting time in our experiment is the number or steps required to measure the TM, equal to 4N. Thus, we can easily increase $M$ by increasing the size of the image recorded without increasing the measurement time. More than just modifying the $N_{DOF}$, the ratio $\gamma = M/N \geq 1$ is expected to change the statistics of the TM. Random Matrix Theory (RMT) predicts that for those matrices the smallest normalized singular value reads $\lambda^0_{\gamma} = (1-\sqrt{1/\gamma})$ ~\cite{marcenko1967distribution, sprik2008eigenvalue}. If we increase $\gamma$ so does the minimum singular value $\lambda^0_{\gamma}$. In a simple physical picture, recording more information at the output results in picking between all available channels those that convey more energy through the medium. If the energy transported by the most inefficient channel reaches and exceeds the noise level, the TM recording is barely sensitive to the experimental noise. We expect that for an appropriate ratio $\gamma$, $\lambda^0_{\gamma}$ reaches the experimental noise level. At this point, no singular values can be drowned in the noise and the pseudo-inverse operator can be efficiently used.

\begin{figure}[ht]
\center
\includegraphics[width=0.5\textwidth]{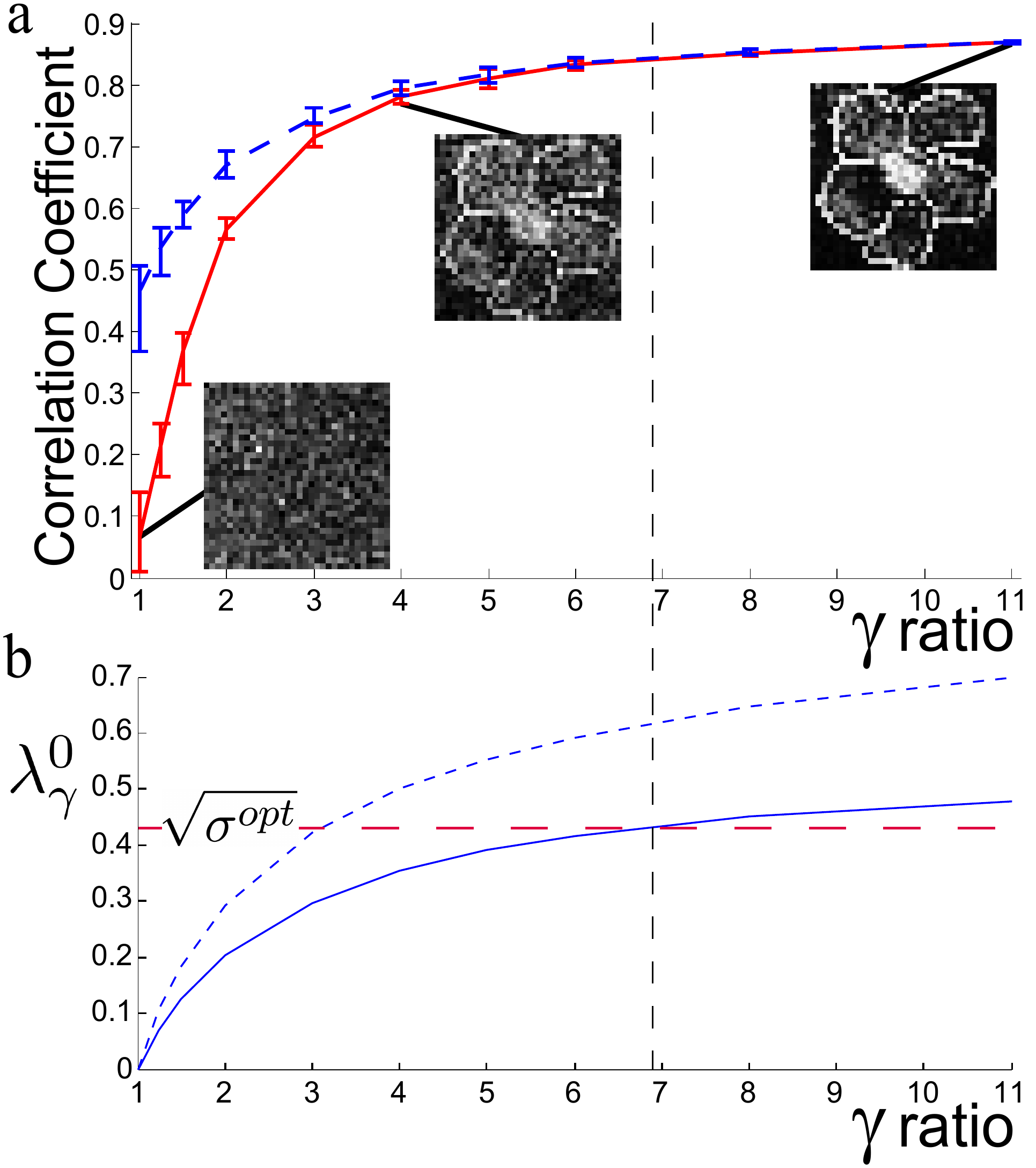}
\caption{   \textbf{Influence of the number of output detection modes. }\textbf{a.} Correlation coefficient between $E_{img}$ and $E_{obj}$ as a function of the asymmetric ratio $\gamma = M/N$ of output to input pixels for MSO (dashed line) and for pseudo-inversion (solid line), without any averaging. Error bars correspond to the dispersion of the results over 10 realizations. \textbf{b.} Experimental (solid line) and Marcenko-Pastur~\cite{marcenko1967distribution} predictions (dashed line) for the minimum normalized singular value as function of $\gamma$. The horizontal line show the experimental noise level $\sqrt{\sigma^{opt}}$.} 
\label{Gamma}
\end{figure}

We experimentally recorded the TM for different values of $\gamma \geq 1$ and tested optimal MSO and pseudo-inversion. The results are shown in Figure \ref{Gamma}.
 Adding degrees of freedom strongly improves the quality of the reconstruction. We see that the quality of the reconstructed image increases with $\gamma$ and reaches a $>85\%$ fidelity for the largest value of $\gamma=11$, without any averaging. The minimum singular value $\lambda^0_{\gamma}$ also increases with $\gamma$. As expected, for $\lambda^0_{\gamma} \gtrsim \sigma^{opt}$, pseudo-inversion is equivalent to optimal MSO. One notice that experimental $\lambda^0_{\gamma}$ are always smaller than their theoretical predictions. This deviation can be explained by the amplitude of the reference pattern $|S_{ref}|$ that induces correlations in the matrix. It is well known in RMT that correlations modify the SVD of a matrix of identically distributed elements~\cite{marcenko1967distribution}.

So far, we tested image transmission in the case of a homogeneous medium, but what would be the results in more complex conditions ? Here, we study the robustness of this technique in presence of ballistic contributions, that is, a fraction of light that have not been scattered at all. The singular values of $K_{obs}$ are proportional to the amplitude transmitted through each channel of the system. Ballistic contributions should give rise to strong singular values corresponding to the apparition of channels of high transmission.  These are not spatially homogeneously distributed in energy, contrarily to multiply scattered contributions. Phase conjugation maximizes energy transmission in channel of maximum transmission~\cite{tanter2000time}. Therefore, ballistic high singular values contributions should be predominant in phase conjugation, independently the image $E_{obj}$ and will not efficiently contribute to image reconstruction. MSO should not be affected since it reaches the optimum intermediate between inversion which is stable except for singular values below noise level, and phase conjugation which forces energy in maximum singular value channels. In other words, MSO will lower weight of channels which do not efficiently contribute to the image reconstruction.

\begin{figure}[ht]
\center
\includegraphics[width=0.5\textwidth]{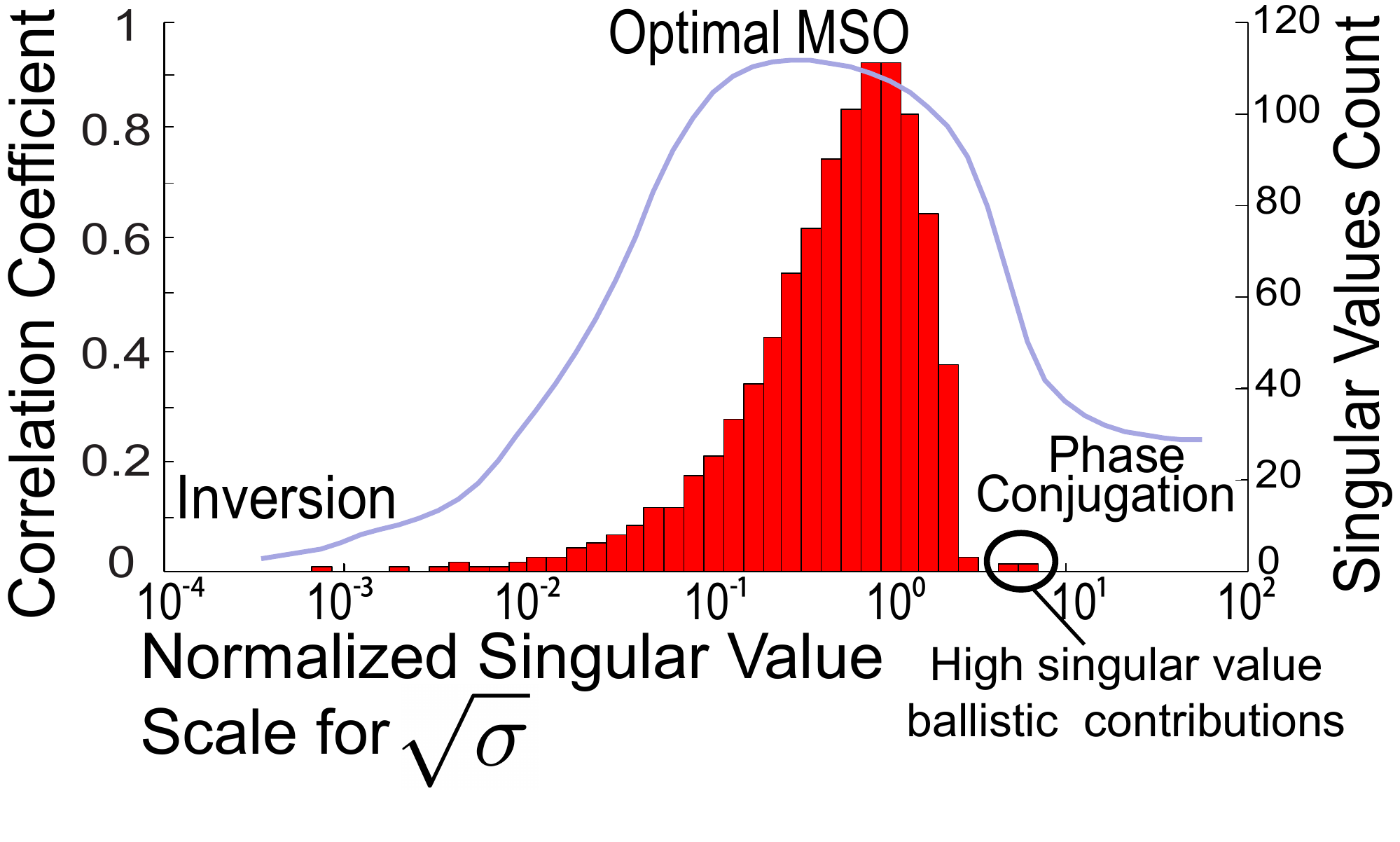}
\caption{\textbf{Influence of the transmission channels on the reconstruction} Correlation coefficient between $E_{img}$ and $E_{obj}$ as a function of $\sqrt{\sigma}$ (line) and singular value distribution of $K_{obs}$ with ballistic contributions in the transmission matrix, averaged over 100 'virtual realizations' of disorder. Both $\sqrt{\sigma}$ and singular values share the same scale on the abscissa axis. Ballistic contributions strongly degrade the reconstruction in phase conjugation while MSO is unaffected.} 
\label{Curves}
\end{figure}

To experimentally study this effect, we moved the collection objective closer to the sample on a thinner and less homogeneous region, where some ballistic light could be recorded. We study in Figure 4 
the quality of the reconstruction as function of $\sigma$ for both experimental conditions (with and without ballistic contributions). Both experiments give comparable results with 93.6\% and 94.5\% correlation coefficient with the optimal MSO operator and both give very low correlation results for inverse matrix operator. With the phase conjugation operator (equivalent to MSO for high $\sigma$), the experiment sensitive to ballistic contributions, give a low correlation coefficient around 35\%, to be compared to the value of 75.7\% we obtained through the multiple scattering sample. This difference can be explained by the presence of a few high singular values contributions (two times greater than the maximum of the other singular values) that perturbate the image reconstruction.

To conclude, we have shown that the transmission matrix allows a rapid and accurate reconstruction of an arbitrary image after propagation through a strongly scattering medium (see supplementary movie). Our approach gives a general framework for coherent imaging in complex media, going well beyond focusing and phase conjugation. It is valid for any linear complex media, and could be extended to several novel photonic materials, whatever the amount of scattering or disorder (from complete disorder to weakly disordered photonic crystals~\cite{garcia2009strong}, and from superdiffusive~\cite{barthelemy2008levy} to Anderson localization~\cite{wiersma1997localization}). The quality of the reconstruction can be increased by harnessing the degrees of freedom of our system, and is very resilient to noise. In addition to its obvious interest for imaging, this experiment strikingly shows that manipulation of wave in complex media is far from limited to single or multi-point focusing. In particular, due to spatial reciprocity, a similar experiment could be performed using an amplitude and phase modulator by shaping the input wavefront to form an image at the output of an opaque medium, which would allow a resolution solely limited by the numerical aperture of the scattering medium~\cite{vellekoop2010exploiting}. The main current limitation is the speed of the TM measurement, which is limited only by the spatial light modulator. Nevertheless faster technologies emerge, such as micromirror arrays or ferromagnetic SLMs, that might in the future widen the range of application domains for this approach, including the  field of biological imaging.

\section*{Methods}

\subsection*{Imaging Setup}
The experimental setup consists on an incident light from a 532 nm laser source (Laser Quantum Torus) that is expanded, spatially modulated by a Spatial Light Modulator (Holoeye LC-R 2500), and focused on an opaque strongly scattering medium  : 80 $\pm$ 25 $\mu m$ thick deposit of ZnO (Sigma-Aldrich 96479) with a measured transport mean free path of $6\pm2$ microns on a standard microscope glass slide. Polarization optics select an almost phase-only modulation mode~\cite{davis2002phasor} of the incident beam, with less than $10\%$ residual amplitude modulation. The surface of the SLM is imaged on the pupil of a 10x objective, thus a pixel of the SLM matches a wave vector at the entrance of the scattering medium. The beam is focused at one side of the sample and the output intensity speckle is imaged on the far side (0.3 mm from the surface of the sample) by a 40x objective onto a 10-bit CCD camera (AVT Dolphin F-145B).

\subsection*{Generation of the amplitude object}
Since there is no simple way to control the amplitude and phase of the incident beam, we generate a virtual amplitude object ($E^{obj} \quad \text{with} \quad s^{obj}_m \in \left[0,1\right]$) by substracting two phase objects. This method is more flexible than placing a real amplitude object in the plane of the SLM. From any phase mask $E^{(1)}_{phase}$ we could generate a second mask $E^{(2)}_{phase}$ where the phase of the $m^{th}$ pixel is shifted by $s^{obj}_m.\pi$. We have $e^{(2)}_m = e^{(1)}_m.e^{i s^{obj}_m\pi}$ with $e^{(j)}_m$ the $j^{th}$ element of $E^{(j)}_{phase}$. $|E^{(2)}_{phase} - E^{(1)}_{phase}|$ is proportional to  $sin(E_{obj}\pi/2)$ and can be estimated by $E_{img} = |W(E^{(2)}_{out} - E^{(1)}_{out})|$ where $E^{(1)}_{out}$ (resp. $E^{(2)}_{out}$) is the complex amplitude of the output speckle resulting from the input vector $E^{(1)}_{phase}$ (resp.  $E^{(2)}_{phase}$). 

\section*{Acknowledgements}

This work was made possible by financial support from "Direction G\'en\'erale de l'Armement" (DGA) and BQR funding from Universit\'e Pierre et Marie Curie and ESPCI. We thank Mathieu Leclerc for performing the sample characterization.


\begin{thebibliography}{30}

\bibitem{goodman2005introduction}
Goodman~J.W.,
\newblock {\em {Introduction to Fourier optics}}.
\newblock (Roberts \& Company Publishers, 2005).

\bibitem{sebbah2001introduction}
Sebbah~P.,
\newblock {\em {Waves and Imaging through Complex Media}}.
\newblock (Kluwer Academic, 2001).

\bibitem{dylov2010nonlinear}
Dylov~D.V. and Fleischer~J.W., 
\newblock {\em {Nonlinear self-filtering of noisy images via dynamical stochastic	resonance}}.
\newblock {\em Nature Photonics}~\textbf{4},~323-328~(2010).

\bibitem{wang1991ballistic}
Wang~L., Ho~P.P., Liu~C., Zhang~G., Alfano~R.R.,
\newblock {\em {Ballistic 2-D imaging through scattering walls using an ultrafast	optical Kerr gate}}.
\newblock {\em Science}~\textbf{253},~769-771~(1991).

\bibitem{shiratori1998photorefractive}
Shiratori~A., Obara~M.,
\newblock {\em {Photorefractive coherence-gated interferometry}}.
\newblock {\em Rev. Sci. Instr.}~\textbf{69},~3741-3745~(1998).

\bibitem{fink1995robust}
Derode~A., Roux~P. and Fink~M.,
\newblock {\em {Robust Acoustic Time Reversal with High-Order Multiple Scattering}}.
\newblock {\em Phys. Rev. Lett.}~\textbf{75},~4206-4209~(1995).

\bibitem{yaqoob2008optical}
Yaqoob~Z., Psaltis~D., Feld~M.S. and Yang~C.,
\newblock {\em {Optical phase conjugation for turbidity suppression in biological	samples}}.
\newblock {\em Nature Photonics}~\textbf{2},~110-115~(2008).

\bibitem{vellekoop2007focusing}
Vellekoop~I.M. and Mosk~A.P.,
\newblock {\em {Focusing coherent light through opaque strongly scattering media}}
\newblock {\em Opt. Lett.}~\textbf{32},~2309-2311~(2007).

\bibitem{cizmar2010situ}
Cizm{\'a}r~T., Mazilu~M. and Dholakia~K.,
\newblock {\em {In situ wavefront correction and its application to micromanipulation}}.
\newblock {\em Nature Photonics}~\textbf{4},~388-394~(2010).

\bibitem{vellekoop2010exploiting}
Vellekoop~I.M., Lagendijk~A. and Mosk~A.P.,
\newblock {\em {Exploiting disorder for perfect focusing}}.
\newblock {\em Nature Photonics}~\textbf{4},~320-322~(2010).

\bibitem{sassaroli1998monte}
Sassaroli~A., Blumetti~C., Martelli~F., Alianelli~L., Contini~D., Ismaelli~A. and Zaccanti~G.,
\newblock {\em {Monte Carlo procedure for investigating light propagation and imaging	of highly scattering media}}.
\newblock {\em Appl. Opt.}~\textbf{37},~7392-7400~(1998).

\bibitem{tanter2000time}
Tanter~M., Thomas~J.L. and Fink~M.,
\newblock {\em {Time reversal and the inverse filter}}.
\newblock {\em J. Acoust. Soc. Am.}~\textbf{108},~223-234~(2000).

\bibitem{montaldo2004}
Montaldo~G., Tanter~M. and Fink~M.,
\newblock {\em {Real time inverse filter focusing through iterative time reversal}}.
\newblock {\em J. Acoust. Soc. Am.}~\textbf{115},~768-775~(2004).

\bibitem{lerosey2007focusing}
Lerosey~G., De~Rosny~J., Tourin~A. and Fink~M.,
\newblock {\em {Focusing beyond the diffraction limit with far-field time reversal}}.
\newblock {\em Science}~\textbf{315},~1120-1122~(2007).

\bibitem{popoff2009measuring}
Popoff~S.M., Lerosey~G., Carminati~R., Fink~M., Boccara~A.C., and Gigan~S.,
\newblock {\em {Measuring the Transmission Matrix in Optics: An Approach to the	Study and Control of Light Propagation in Disordered Media}}.
\newblock {\em Phys. Rev. Lett.}~\textbf{104},~100601~(2010).  

\bibitem{marcenko1967distribution}
Mar{\v{c}}enko~V., Pastur~L.,
\newblock {\em {Distribution of eigenvalues for some sets of random matrices}}.
\newblock {\em Sbornik: Mathematics}~\textbf{1},~457~(1967)  

\bibitem{introPaulraj}
Gore~D., Paulraj~A. and Nabar~R.,
\newblock {\em Introduction to Space-Time Wireless Communication}.
\newblock (Cambridge University Press,~2004).


\bibitem{arridge1999optical}
Arridge~S.R.,
\newblock {Optical tomography in medical imaging}.
\newblock {\em Inverse problems}~\textbf{15},~R41-R94~(1999).

\bibitem{corlu2007three}
Corlu~A., Choe~R., Durduran~T., Rosen~M.A., Schweiger~M., Arridge~S.R., Schnall~M.D., and Yodh~A.G., 
\newblock {\em {Three-dimensional in vivo fluorescence diffuse optical tomography	of breast cancer in humans}}.
\newblock {\em Optics Express}~\textbf{15},~6696-6716~(2007).


\bibitem{maire2009experimental}
Maire~G., Drsek~F., Girard~J., Giovannini~H., Talneau~A., Konan~D., Belkebir~K., Chaumet~P.C., Sentenac~A., 
\newblock {\em {Experimental demonstration of quantitative imaging beyond Abbe's limit with optical diffraction tomography}}.
\newblock {\em Phys. Rev. Lett.}~\textbf{21},~213905~(2009).


\bibitem{yariv1979compensation}
Yariv~A., Fekete~D. and Pepper~D.M.,
\newblock {\em {Compensation for channel dispersion by nonlinear optical phase conjugation}}.
\newblock {\em Optics Letters}~\textbf{4},~52~(1979).

\bibitem{derode2001scattering}
Derode~A., Tourin~A. and Fink~M.,
\newblock {\em {Random multiple scattering of ultrasound. II. Is time reversal a self-averaging process?}}.
\newblock {\em Phys. Rev. E}, ~\textbf{64},~036606~(2001).

\bibitem{sprik2008eigenvalue}
Sprik~R., Tourin~A., De~Rosny~J. and Fink~M.,
\newblock {\em {Eigenvalue distributions of correlated multichannel transfer matrices in strongly scattering systems}}.
\newblock {\em Phys. Rev. B}~\textbf{78},~12202~(2008) 

\bibitem{tikhonov1963solution}
Tikhonov~A.N.,
\newblock {\em {Solution of incorrectly formulated problems and the regularization	method}}.
\newblock {\em Soviet Math. Dokl}, ~\textbf{4},~1035~(1963).
 
\bibitem{lemoult2009}
Lemoult~F., Lerosey~G., De~Rosny~J. and Fink~M.,
\newblock {\em {Manipulating Spatiotemporal Degrees of Freedom of Waves in Random	Media}}.
\newblock {\em Phys. Rev. Lett.}~\textbf{103},~173902~(2009).  

\bibitem{davis2002phasor}
Davis~J.A., Nicol{\'a}s~J. and M{\'a}rquez~M.A.,
\newblock {\em {Phasor analysis of eigenvectors generated in liquid-crystal displays}}.
\newblock {\em Appl. Opt.}~\textbf{41},~4579-4584~(2002).

\bibitem{garcia2009strong}
Garc\'\i{}a~P.D., Sapienza~R., Froufe-P\'erez~L.S. and L\'opez~C.,
\newblock {\em  {Strong dispersive effects in the light-scattering mean free path in photonic gaps}}.
\newblock {\em Phys. Rev. B}~\textbf{79},~241109~(2009)  

\bibitem{barthelemy2008levy}
Barthelemy~P., Bertolotti~J. and Wiersma~D.S.,
\newblock {\em {A L{\'e}vy flight for light}}.
\newblock {\em Nature}~\textbf{453},~495-498~(2008)  

\bibitem{wiersma1997localization}
Wiersma~D.S., Bartolini~P., Lagendijk~A. and Righini~R.,
\newblock {\em {Localization of light in a disordered medium}}.
\newblock {\em Nature}~\textbf{390},~671-673~(1997).


\end{thebibliography}
\end{document}